\documentclass[11pt]{article} 

\usepackage[utf8]{inputenc} 

\usepackage{amsmath,amsfonts,amssymb}
\newcommand{\qed}{\hfill $\Box$}
\usepackage{geometry} 
\usepackage{graphicx}
\usepackage{epstopdf}
\usepackage{booktabs} 
\usepackage{array} 
\usepackage{paralist} 
\usepackage{verbatim} 
\usepackage{subfig} 

\usepackage{ulem}

\usepackage{fancyhdr} 
\usepackage{sectsty}
\allsectionsfont{\sffamily\mdseries\upshape} 

\usepackage[nottoc,notlof,notlot]{tocbibind} 
\usepackage[titles,subfigure]{tocloft} 

\newcommand{\field}[1]{\mathbb{#1}}
\newcommand {\R}{\field{R} }

\newcommand {\C} {\field{C} }

\newcommand {\eps}{\varepsilon }

\newcommand {\ol}{\overline }

\newtheorem{prop}{Proposition}[section]

\newtheorem{rem}[prop]{Remark}

\newtheorem{theorem}[prop]{Theorem}
\newtheorem{lemma}[prop]{Lemma}
\newtheorem{cor}[prop]{Corollary}

\usepackage{color}
\usepackage{colordvi}
\definecolor{magenta}{rgb}{.5,0,.5}
\definecolor{black}{rgb}{1.0,1.0,1.0}
\definecolor{magenta}{rgb}{.1,0,.3}
\definecolor{gruen}{rgb}{0.2,0.5,.5}
\definecolor{light}{rgb}{ 0.992, 0.961,  0.902}
\definecolor{Tan}{rgb}{ 0.992, 0.9,  0.902}

\newcommand{\komment}[1]{{}}

\title{Age structure, replicator equation, and the prisoner's dilemma}
\author{Sona John\footnote{Department of Mathematics, Technical University of Munich, 85748 Garching, Germany}, Johannes M\"uller\footnote{Department of Mathematics, Technical University of Munich, 85748 Garching, Germany and Institute for Computational Biology, Helmholtz Center Munich, 85764 Neuherberg, Germany
}}

\begin{document}
\maketitle

\begin{abstract}
We investigate the evolutionary dynamics of an age-structured population under 
weak frequency-dependent selection. It turns out that the weak selection is affected in a non-trivial way by the life-history trait. We can disentangle the dynamics, based 
on the appearance of different time scales. These time scales, which seem to form a universal structure in the interplay of weak selection and life-history 
traits, allow us to reduce the infinite dimensional model to a one-dimensional modified replicator equation. The modified replicator equation is then used to investigate cooperation (the prisoner's dilemma) by means of adaptive dynamics. We identify conditions under which age structure is able to promote cooperation. At the end we discuss the relevance of our findings.
\end{abstract}

\par\bigskip

\section{Introduction}

As we all know, every individual is unique, we all have our individual biography, have a certain age, are in a specific situation, circumstances which strongly influence our response to selective pressure. Most evolutionary studies, however, treat individuals as identical and  featureless  particles. This approach, deeply rooted in statistical physics, allows to formulate and analyze powerful models. In those models, individuals neither have a history nor do they differ from fellow individuals of the same species, which is biologically unrealistic. In population genetics, perhaps the Kingman coalescence~\cite{durrett:popGen} and the replicator equation~\cite{Hofbauer2003} are the most important and prominent representatives of this approach. Realistic models need to take  life history traits into account. Herein,  ``life history'' covers all factors affecting survival and reproduction, as age, sex ratio, dormancy, plasticity, cannibalism or dispersal to name but a few~\cite{Dingle1982,Roff1993,Lennon2021}. In recent years, the evolutionary theory advanced to incorporate life-history traits  into evolutionary models, but it still remains a challenge to reasonably set up and analyze such models. As we will see, one major point is simply technical: Inevitably, the state space of these models blow up. 
\par\medskip

The assumption that all individuals have the same fertility rate through out its life span is very strong and biologically unrealistic. More appropriate approach is to subdivide the population into different age classes to include the population structure ~\cite{Gotelli2001}. The classical way of modeling age structured population is similar to Bernadelli-Lewis-Leslie matrices~\cite{Bernadelli1941,Lewis1942,Leslie1945}. However, as age is not a discrete state, continuous time models may be even better suited, as it is done in  the  Sharpe–Lotka–Feller renewal model~\cite{Lotka1939,Feller1941}. The renewal equation is equivalent to the hyperbolic age structured partial differential equation (PDE) models such as the one proposed by McKendrick~\cite{McKendrick1926}, that can also be obtained by means of a continuum limit from matrix models mentioned above. As long as nonlinear interactions do not play a role, these models are very well understood~\cite{Iannelli2017,Inaba2017}. Nonlinearity bear more challenges~\cite{Webb1985,Magal2018}.\\
In respect to evolutionary theory, a large body of literature about age as a life-history traits investigates the inverse problem~\cite{Kozlowski1987}: How should I design my life-history trait to optimize fitness? This fruitful approach helps to explain numerous observations, such as  delayed maturation~\cite{Calsina2000}, conditions inducing  semelparity and iteroparity~\cite{Oizumi2014}, or the size-at-age of fish~\cite{Vincenzi2014} to name but a few. 
\par\medskip

The game theoretical approach in evolutionary theory mainly focus on the impact of intra-species (or also inter-species) interactions on the evolution of each species~\cite{HofbauerSigmund:book, Hofbauer2003}. The impact of interactions are translated to the growth rate of each species via the fertility and mortality factors. This provides the replicator equation which can describe the growth rate of each of these types based on the interaction with other types. Particularly the question how (bacterial) cooperative behavior can be understood is addressed by this approach~\cite{West2006,Nowak2006,Hoesel2020}. 
One obvious short coming of these models is that they largely ignore the underlying structure of different life history traits,such as dormancy, seed banks or age. Already McNamara~\cite{McNamara2013} indicated that game theory should be enriched by the incorporation of life-history traits. However, in that often the dimensions of the models explode, as can be seen in the article~\cite{Li2015}, where a game theoretical model is augmented by a discrete age structure of the Leslie-type -- (continuous) age structure even leads to infinite dimensional models. If we reformulate this observation in biological terms, populations have a higher degree of freedom to respond to evolutionary forces. We need to deal with this rich variety of potential reactions.\\


How to include the population structure based on a life history trait such as age in a population dynamics generated by the replicator equations is still not completely understood~\cite{Caswell2011}. An early attempt to  combine  these two approaches is the paper by Garay et. al, 2016 ~\cite{Garay2016} on sib cannibalism. This approach show that the only way to explain the evolution of a strategy such as sib cannibalism is by taking into account the age structure of the populations. Incorporating the age structure into a matrix game ~\cite{Li2015,Lessard2018}, results in increasingly complicated analysis with high dimensionality of matrix. Numerous  articles~\cite{Wu2009,Szolnoki2009,Rong2010,Liu2012,Rong2013,Souza2020} avoid all these  technicalities and simply use individual based simulation models. They find that cooperation can be stabilized by age structure, but the insight by simulation-based studies is, of course, limited. 
Recent study in this direction by Argasinski and Broom~\cite{Argasinski2021} try to overcome the challenges of a matrix game approach by developing a high dimensional ODE system to describe the age structure. The generalized ODE system with continuous time but discrete age classes is obtained by application of delayed differential equations~\cite{Caswell2001}.  When this combined with the multi population evolutionary game, they obtain a mixed PDE-ODE system which is simpler to handle mathematically. This model is applied to understand the impact of age specific mortality on observed sex ratios. \\

The present paper follows the ideas introduced by Argasinski and Broom~\cite{Argasinski2021}, but simplifies the formalism on the one hand as we base our model on a time-continuous age structure, on the other hand as we focus on weak selection. Weak selection leads to time scales differences, which we utilize to reduce the dimension of the system drastically. 
We obtain a modified replicator equation by life-history related lumped parameters that capture the impact of age structure. These parameters also holds a clear biological interpretation. Thus,
we keep the 
structure of the replicator equation as a fundamental and simple model, but show how to adapt the equation to take life-history traits into account. The resulting modified replicator is used as a basis for adaptive dynamics to investigate the stability of cooperation. We feel that this approach adds value to the numerous simulation models which show that age structure can promote cooperation.\par\medskip 

The paper is structured as follows: We first (section~2) review the replicator equation  under the assumption of weak frequency selection. The main features we find here are then used in the analysis of age-structured models. In section~3, we recall well known results from the theory of age structured models, and reformulate the standard
models in a way suited for our task. The main work is done in section~4, where we introduce frequency-dependent selection into the age-structured models, and particularly show under which conditions it is possible to obtain the modified replicator equation, which is used as an ingredient for adaptive dynamics in section~5 to analyze the prisoner's dilemma. At the end, we discuss the relevance of our findings.

\section{Prelude: The replicator equation in case of weak frequency-dependent selection}

We briefly repeat the derivation of the classical replicator equation, which can be found in numerous monographs, see e.g.~\cite{HofbauerSigmund:book, Hofbauer2003}. The aim is to understand the necessary steps we will need to do to include age as a life-history trait in the derivation of the replicator equation. We focus on weak frequency-dependent selection: Quite often, fitness differences induced by a given trait are relatively small. We therefore distinguish between the background fitness of an organism, which is $\mathcal O(1)$, and the fitness induced by a given, frequency-dependent trait, which modifies the background fitness of an individual by $\mathcal O(\eps)$. \\
We consider types $i=1,..,n$ with population size $X_i$. The frequencies are denoted by $x_i=X_i/(\sum_j X_j)$. The growth of type $i$ is described by 
$$ \dot X_i = (f_i+\eps g_i(x))\, X_i$$
where $f_i$ addresses the background fitness of species $i$, $\eps g_i(x)$ models the weak effect frequency-dependent trait on the fitness, and  $x=(x_1,..,x_n)$. The r.h.s. is homogeneous of degree one, and hence we are able to derive proper equations for the frequencies $x_i$, 
\begin{eqnarray}
\dot x_i = \frac{\dot X_i}{\sum_j X_j}-\frac{\sum_j\dot X_j}{\sum_jX_j}\,\frac{X_i}{\sum_j X_j}
= \bigg( f_i- \sum_j x_jf_j\bigg) x_i + \eps\,\bigg( g_i(x)- \sum_j x_jg_j(x)\bigg) x_i. \label{rpliFist}
\end{eqnarray}
By the way, the total population size $N=\sum_j X_j$ also satisfies a proper equation,
$$ \dot N =  \bigg(\sum_j x_j(f_j+\eps\,g_j(x))\bigg)\, N.$$
That is, we can disentangle the dynamics of the frequencies and the dynamics of the total population size. Together, with $x$ and $N$, we are able to recover the original variables $X_i=x_i N$.\\
Now we assume that the background fitness $f_j$ is equal for all types, $f_j=f$; otherwise, the weak frequency-dependent fitness differences will not affect the dynamics crucially. Then we obtain the well known replicator equation
\begin{eqnarray}
	\dot x_i &=&  \eps\, \bigg( g_i(x)- \sum_j x_jg_j(x)\bigg) x_i.
\end{eqnarray}
As we have seen, we used two main ingredients: \\
(a) The background fitness $f_i$ is identical, and the frequency-dependent fitness differences are weak. \\
(b) The equation for the population size is homogeneous of degree one.\\
Assumption (b) allows to disentangle to dynamics for the total population size and for relative frequencies. Assumption (a) implies that the frequencies of types, driven by weak selection,  only change on a slow time scale.\par\medskip

If we add age structure as an life-history trait, we will use a similar route for the analysis as we did above. That is, we first disentangle total population size and (age structured) relative frequencies, which of course is more involving than in the unstructured case. We also will use the second observation we just discussed, that weak selection drives the system only on a slow time scale: Therefore, there is sufficient time for the life-history trait time to evolve under fairly constant conditions. The life-history trait tends to its (quasi) equilibrium: The age distributions of the subpopulations are well approximated by their equilibrium distributions. In this, the relative frequency of a type is sufficient to characterize its full age-structured relative abundance. We are able to reduce the dimension of the system, and return to an ODE for the description of the long term dynamics.

\section{Age structure -- linear model}
In this section, we basically recall well known facts about age structured populations, introduce notation, and re-arrange the mathematical structure in a convenient way. Nice introductions into the theorems and concepts about age structured models used here can be e.g.\ found in the monographs~\cite{Webb1985,Perthame2006,Iannelli2017,Inaba2017}. In the present section, we particularly follow the book by Inaba~\cite{Inaba2017}. For the convenience of the reader, and since the linear theory is the basis for the nonlinear theory develop below, we repeat also well known results.

\subsection{Model}
We consider two independent age structured populations with density $u_1(a,t)$ for the first, and $u_2(a,t)$ for the second population, described by McKendrick-von Foerster equation as
\begin{eqnarray}
(\partial_t+\partial_a)u_i(a,t) &=& -\mu_i(a) \, u_i(a,t)\\
u(0,t) &=& \int_0^\infty \beta_i(a)\, u_i(a,t)\, da.
\end{eqnarray}
At this stage, it does not seem to make sense to have two populations, but later we will have them to interact. In order to have a consistent notation, we right now start with two populations. 
Let us assume that the birth rates $\beta_i(a)$ have compact support. The rates are non-negative, the birth rate is not identically zero, and the rate functions are in $C^0$. \\
 
Standard theory indicates that $u_i(a,t)$ will, in the long run, grow exponentially, such that (in a sense that we discuss more in detail below) eventually $u_i(a,t)=c_i\,\varphi(a)\,e^{\lambda_i\, t}$, where $\lambda_i\in\R$ and $\varphi_i\geq$ 0 are uniquely determined by the fact that $\varphi_i$ is the normalized, leading (Perron) eigenfunction. As we know $u_i(a,t)$ give the fraction of individuals with age $a$ at time $t$, which can be related to the probability of an individual born at time $t-a$ and survive till an age $a$ given as $l_i(a)=e^{-\int_0^a\mu_i(s)\, ds}$. With the assumtion of exponential growth we can write $u_i(a,t) = c_i e^{\lambda_i(t-a)} l(a) = c_i \varphi_i(a) e^{\lambda_i t}$.

If we plug this expression into the McKendirck equation, and introduce the probability $l_i(a) = e^{-\int_0^a\mu_i(s)\, ds}$ for an individual to survive up to age $a$, the characteristic equation for the eigenvalue $\lambda_i$ reads
$$ 1 = \int_0^\infty \, e^{-\lambda_i a} l_i(a) \,\beta_i(a)\, da$$
while the eigenfunction $\varphi_i(a)$ is given by 
$$ \varphi_i(a)= b_i \,e^{-\lambda_i a} l_i(a),$$
where we choose $b_i$ such that $\|\varphi\|_{L^1}=1$,
$b_i = \left(\int_0^\infty e^{-\lambda_i a} l_i(a)\, da\right)^{-1}$. For technical reasons, we assume that $\mu_i(a)$ are asymptotically large enough to ensure that not only $\varphi_i(a)\in L^1$, but even
\begin{eqnarray}\label{expoAssumption}
\exists \delta >0 : \varphi_i(a)\, e^{\delta a}\in L^1(\R_+).
\end{eqnarray}
For biological relevant parameters, this assumption is not really a restriction, as the mortality in high age classes mostly becomes arbitrary large and $l(a)$ tends to zero in the large age limit.

We furthermore assume neutrality in the competition of both populations (in the replicator equation, this assumption was that $f_j$ are the same for all types), that is, we require
$$ \lambda_1 = \lambda_2.$$
We define $\lambda:= \lambda_1 = \lambda_2$. Note that this condition in general differs from the condition that the reproduction numbers $R_{0,i} =  \int_0^\infty \, l_i(a)\,\beta_i(a)\, da$ are identical. We do not address populations in equilibrium, but exponentially growing populations, and hence do not focus on the reproduction number but on the exponential growth to establish neutrality. We conjecture that a parallel theory investigating a constant population size requires that both population have the identical reproduction number, cf~\cite{Sellinger2019} and \cite{Mueller2022}. \par\medskip 
Using operator notation, we re-write the model in a handy way,
\begin{eqnarray}
 \partial_tu_i = A_i u
\end{eqnarray}
where $A_i:D(A_i)\subset L^1\rightarrow L^1$, 
$$ A_i u(a) = - \partial_a u(a) - \mu_i(a) u(a)$$
and
$$ D(A_i) = \{u\in L^1\,|\, u(0) = \int_0^{\ol a} \beta_i(a) u(a)\, da,\,\,\, A_iu\in L^1\}. $$
We note a few well known but non-trivial facts about the spectrum of $A_i$. We already have one eigenvalue $\lambda_i$, which is simple, together with the corresponding, positive eigenfunction $\varphi_i$. We formulate these facts for $A_1$, but clearly, the theorem also holds for $A_2$. Find the proof of the following theorem in~\cite[Proposition 10.3]{Inaba2017}.

\begin{theorem} (a) If $\mbox{supp}(\beta_1)$ is compact, the spectrum of $A_1$ only consist of point spectrum. (b) The complex half-plane $\{z\in\C\,|\,\Re(z)>\sigma\}$ for $\sigma\in\R$ only contains at most a finite number of eigenvalues. (c) Any eigenvalue $\tilde\lambda$ of $A_1$ with $\tilde\lambda\not = \lambda_1$ has a real part smaller than $\lambda_1$, $\Re(\tilde\lambda)<\lambda_1$.
\end{theorem}

From (b) and (c) we conclude that we have a spectral gap: There is $\delta>0$, such that the maximal real part of all eigenvalues of $A_i$ unequal $\lambda_i$ is at most $\lambda_i-\delta$.\par\bigskip

\subsection{Ajoint operator}
Next we introduce the adjoint operator. Note that we do not aim at deep functional analytic results about operators and their adjoints; for us, the adjoint operator rather is a convenient notational shortcut for integration by parts and rearranging terms. Therefore we do not really care about function spaces, domain (which we do not define), and alike. A deeper and rigorous discussion of the adjoint operator can be found in~\cite{Perthame2006,Feng2005}.
\begin{prop}
The adjoint operator $A_i^+:D(A_i^+)\subset L_1^*\rightarrow L_1^*$ is given by 
$$A^+_i[\psi](a) =  \partial_a  \psi(a)- \mu_i(a) \psi(a)+\beta_i(a) \psi(0).$$
\end{prop}
{\bf Proof: }
For $u\in D(A_i)$ and a test function $\psi(a)\in C^1$ we have 
\begin{eqnarray*}
\int_0^\infty \psi(a)\bigg(- \partial_a u(a) - \mu_i(a) u(a)\bigg)\, da 
&=& - \psi(a) u(a)\bigg|_{a=0}^{\infty} + \int_0^{\infty}  u(a)\bigg( \partial_a  \psi(a)- \mu_i(a) \psi(a)\bigg)\, da\\
&=&  \int_0^{\infty}  u(a)\bigg( \partial_a  \psi(a)- \mu_i(a) \psi(a)+\beta_i(a) \psi(0)\bigg)\, da
\end{eqnarray*}
and hence $A^+_i[\psi](a) =  \partial_a  \psi(a)- \mu_i(a) \psi(a)+\beta_i(a) \psi(0)$. 
\par\qed\par\medskip

Also the adjoint eigenfunctions (the eigenfunction of the adjoint operator) for $\lambda_i$ and its properties are well known~\cite{Perthame2006}.
\begin{prop}
The adjoint eigenfunctions for $\lambda_i$ read 
\begin{eqnarray}
	\psi_i(a) 
	&=& c_i \bigg(e^{\int_0^a \mu_i(s)\, ds+\lambda_i\, a}
	- \int_0^a e^{\int_b^a \mu_i(s)\, ds+\lambda_i\, (a-b)} \beta_i(b)\, db\bigg)\\
&=&	c_i \int_a^\infty e^{-\lambda_i (s-a)} \frac{l_i(s)}{l_i(a)} \beta_i(s) ds
\end{eqnarray}
\end{prop}

{\bf Proof: }
The adjoint eigenfunctions for $\lambda_i$ satisfy
$ A_i^+\psi_i=\lambda_i\psi_i$, that is,
\begin{eqnarray*}
\psi_i'(a) &=& (\mu_i(a)+\lambda_i)\psi_i(a) - \beta_i(a)\psi_i(0).
\end{eqnarray*}
The variation-of-constants formula give us
$$\psi_i(a) 
= e^{\int_0^a \mu_i(s)\, ds+\lambda_i\, a} \,\psi_i(0)
- \int_0^a e^{\int_b^a \mu_i(s)\, ds+\lambda_i\, (a-b)} \,\beta_i(b)\,\psi_i(0)\, db.
$$
With $c_i=\psi_i(0)$ and the characteristic equation $\int_0^\infty e^{-\lambda_i b} l_i(b)  \beta_i(b) db=1$ we obtain
\begin{eqnarray*}
 \psi_i(a) &=& c_i \frac{e^{\lambda_i a}}{l_i(a)} \bigg( 1 - \int_0^a e^{-\lambda_i b} l_i(b)  \beta_i(b) db \bigg)
 = \ol c_i \frac{e^{\lambda_i a}}{l_i(a)} \bigg(\int_0^\infty e^{-\lambda_i b} l_i(b)  \beta_i(b) db -  \int_0^a e^{-\lambda_i b} l_i(b)  \beta_i(b) db   \bigg) \\ \nonumber
 &=& c_i \frac{e^{\lambda_i a}}{l_i(a)} \bigg(\int_a^\infty e^{-\lambda_i b} l_i(b)  \beta_i(b) db   \bigg).
\end{eqnarray*}
\par\qed\par\medskip

As an immediate result, we note that the support of $\psi_i(a)$ is a determined by the support of $\psi_i(a)$: If $\mbox{supp}(\beta_i)\subset[0,\ol a]$, then $\mbox{supp}(\psi_i)\subset[0,\ol a]$. This is a direct consequence of $\psi_i(a) = \ol c_i \int_a^\infty e^{-\lambda_i (s-a)} \frac{l_i(s)}{l_i(a)} \beta_i(s) ds$.\\
We note furthermore that in the special case of constant $\mu_i$ and $\beta_i$, we find that also $\psi_i(a)=\ol c_i$ are constant; in the straight forward calculation, $\lambda_i=\beta_i-\mu_i$ is used.
\subsection{Appropriate normalization or: the reproductive value}
\label{normal}

Recall that $b_i = \left(\int_0^\infty e^{-\lambda_i a} l_i(a)\, da\right)^{-1}$; additionally, we introduce
$$ T_i = \int_0^\infty a\,\beta_i(a)\,e^{-\lambda_i a} l_i(a)\, da.$$
Then,
$$ \frac{\int_0^\infty \beta_i(a)\,e^{-\lambda_i a} l_i(a)\, da}{\int_0^\infty e^{-\lambda_i a} l_i(a)\, da}
=
b_i\,\int_0^\infty \beta_i(a)\,e^{-\lambda_i a} l_i(a)\, da$$
can be interpreted as the crude birth rate (CBR) of the population, while
$$ \ol A_i := \frac{\int_0^\infty a\,\beta_i(a)\,e^{-\lambda_i a} l_i(a)\, da}{\int_0^\infty \beta_i(a)\,e^{-\lambda_i a} l_i(a)\, da} =
\frac{T_i}{\int_0^\infty \beta_i(a)\,e^{-\lambda_i a} l_i(a)\, da}$$
is the average age at childbearing, booth in the exponentially growing population. The product $b_i\,T_i$ obviously can be seen as the CBR times the average age at childbearing. Now we choose for the normalization of the adjoint eigenfunctions $\psi_i(a)$ that
$$c_i = 1/(b_i\, T_i)$$
such that with the normalization chosen for $\psi_i(a)$ and $\varphi_i(a)$, we obtain (exchange the order of the integrals)
$$ \int_0^\infty \psi_i(a)\,\varphi_i(a)\,da
 =
 \int_0^\infty \frac 1 {b_i\,T_i}
 \int_a^\infty e^{-\lambda_i (s-a)} \frac{l_i(s)}{l_i(a)} \beta_i(s) ds
 \,\,b_i\, e^{-\lambda_i\, a}l_i(a)\, da
 =
\frac{T_i}{T_i}=1.$$
The function $\psi_i(a)$ is also called normalized reproductive value, as $\psi_i(a)$ gives (in a normalized way) the information how much an individual of age $a$ contributes in her future life to the ancestry of future generations~\cite[Sect. 1.3.2]{Inaba2017}.
For our future needs, we note that
\begin{eqnarray} \label{normEverytingTogether}
\psi_i(0) \int_0^\infty\beta_i(a)\,\varphi_i(a)\, da = \frac 1 {\ol A_i}.
\end{eqnarray}

\subsection{Relative frequencies}\label{relativeFrequenciesLinear}
In the replicator equations, we did go from the population size $X_i$ to relative frequencies $x_i$. We do the same for our model, and disentangle the dynamics of population size and relative frequencies (in age structure and type $i=1,2$). It is not convenient to use the total population size $\int_0^\infty u_1(a,t)+u_2(a,t)\, da$ (which parallels $\sum_jX_j$) as a reference. Instead, we use a weighted population size, where the adjoint eigenfunctions are used as weights,
$$N(t) = \int_0^\infty \,\psi_1(a)\,u_1(a,t)+\psi_2(a) u_2(a,t)\,da.$$
As it will become clear below, this choice is closely related to spectral projectors, and thus yields a convenient mathematical structure.

\begin{prop}
$$N(t) = e^{\lambda t } N(0).$$
\end{prop}
{\bf Proof: }
\begin{eqnarray*}
	N'(t) 
	&=& \int_0^\infty \partial_t(\psi_1(a)\,u_1(a,t)+\psi_2(a)\,u_2(a,t))\, da 
	=\int_0^\infty (\psi_1(a)\,A_1[u_1](a,t)+\psi_2(a)\,A_2[u_2](a,t))\, da\\
	&=&\int_0^\infty (A_1^+[\psi_1](a)\,u_1(a,t)+A_2^+[\psi_2](a)\,u_2(a,t))\, da
	=\int_0^\infty \,\lambda_1\psi_1(a)\,u_1(a,t)+\lambda_2\psi_2(a) u_2(a,t)\,da.
\end{eqnarray*}
The result follows with $\lambda=\lambda_1=\lambda_2$.
\par\qed\par\medskip 

From this time on, we assume that $N(0)>0$: A part of the initial population is assumed to be in the fertile age classes. This is not really a restriction, as otherwise the population dies out without producing any offspring. Next we define 
$$\nu_i(a,t) = u_i(a,t)/N(t).$$
Then $\nu_i\in L^1_+(\R_+)$ and 
$$\|\psi_1\nu_1\|_{L^1}+\|\psi_2\nu_2\|_{L^1}= \|\psi_1\nu_1+\psi_2\nu_2\|_{L^1}= \int_0^\infty \psi_1(a) \nu_1(a,t)+\psi_2(a) \nu_2(a,t)\, da = 1.$$
Furthermore, 
\begin{eqnarray*}
 \partial_t \nu_i(a,t) &=& 
\partial_t \frac{u_i(a,t)}{N(t)} 
= \frac{\partial_t u_i(a,t)}{N(t)} -\frac{N'(t)}{N(t)}  \frac{u_i(a,t)}{N(t)} \\
&=& - \partial_a\nu(a,t) - (\mu_i+\lambda)\nu(a,t)
= A_i[\nu_1](a,t)-\lambda \nu_i(a,t)
\end{eqnarray*}
and
$$ \nu_i(0,t) = \frac{u_i(0,t)}{N(t)} = \int_0^\infty \beta_i(a)\,\nu_i(a,t)\, da.$$
That is, $N(t)$ measures the total population in an appropriate way, and is exponentially growing. The variables $\nu_i(a,t)$ incorporate the information about relative frequencies. The original solution $u_i(a,t)$ can be reconstructed from $N(t)$ and $\nu_i(a,t)$. We do not loose information.

\subsection{Spectral projectors}
The subpopulations in the classical replicator equation did not have structure. This is different for the age structured model. However, what we will find out here is that in the long run, the subpopulations tend to a very specific age structure. That is, in the infinite dimensional space, there is a one-dimensional subspace, where the solution will tend to. We aim to reformulate the dynamics in such a way that this behavior becomes explicit.\par\medskip

Let $\Pi_i^+:L^1\rightarrow L^1$ the spectral projector
$$ \Pi_i^+[u] = \int_0^\infty \psi_i(a)\,u(a)\, da\,\varphi_i(a)$$
and $\Pi_i^-:L^1\rightarrow L^1$ given by
$$ \Pi_i^-[u] = (id-\Pi_i^+)[u](a).$$
Obviously, the rank of $\Pi_i^+$ is $1$ (we map into a one-dimensional manifold spanned by $\varphi_i$), and for $u\in L^1_+$, 
$$ \int_0^\infty \psi_i(a)\Pi_i^-[u](a)\, da = 0.$$
Furthermore, $\Pi_i^++\Pi_i^- = id$. \par\medskip 

As we know that $\nu_i(a,t)\in L^1_+$,  $\|\psi_1\nu_1+\psi_2\nu_2\|_{L^1}=1$, and $\|\varphi_i\|_{L^1}=1$,  we have 
$$ \|\Pi_1^+[\nu_1]\|_{L^1} +\|\Pi_2^+[\nu_2]\|_{L^1}  = 1.$$
We only need to follow $\Pi_1^+[\nu_1]$, as from that value, we can construct $\Pi_2^+[\nu_2]$ by
$$ \Pi_2^+[\nu_2](a) = (1-\|\Pi_1^+[\nu_1]\|_{L^1})\,\,\varphi_2(a).$$

Let us return to the dynamics. With
\begin{eqnarray}
	x(t) = \int_0^\infty \psi_1(a) \nu_1(a,t)\, da
\end{eqnarray}
we have
$ \Pi_1 ^+[\nu(.,t)] = \int_0^\infty\psi_1(a) \nu_1(a,t)\, da\,\varphi_i(a) 
= x(t)\, \varphi_1(a)$. 
If we additionally define 
\begin{eqnarray}
\eta_i(a,t) = \Pi^-[\nu_i](a,t),
\end{eqnarray}
then $x(t)$ and $\eta_i(a,t)$ characterize completely $\nu_i(a,t)$ (we can reconstruct $\nu_i(a,t)$ by $x(t)$ and $\eta_i(t)$)
\begin{eqnarray*}
\nu_1(a,t) &=& 
\Pi^+_1[\nu_1](a,t)+\Pi^-_1[\nu](a,t)
=
x(t)\, \varphi_1(a) + \eta_1(a,t)\\
\nu_2(a,t) &=& 
\Pi^+_2[\nu_2](a,t)+\Pi^-_2[\nu](a,t)
=
(1-x(t))\, \varphi_2(a) + \eta_2(a,t)
\end{eqnarray*}
Basically, we did construct a new coordinate system, where the exponentially growing component of $(u_1(a,t),u_2(a,t))$ is mapped to $(x(t)\varphi_1(a), (1-x(t))\varphi_2(a))$, and the functions $\eta_i(a,t)$ measure the difference of the age structure $\nu_i(a,t)$ from the age structure given by the exponentially growing solution $\varphi_i(a)$.

\begin{theorem}
\begin{eqnarray}
\dot x &=& 0\\
\partial_t\eta_i &=& A_i[\eta_i]-\lambda \eta_i,\qquad 
\eta_i(0,t)= \int_0^\infty \beta_i(a)\eta_i(a,t)\, da,\qquad i=1,2.
\end{eqnarray}
\end{theorem}
{\bf Proof: } 
\begin{eqnarray*}
	\frac d {dt} x(t)
	&=& \int_0^\infty \psi_1(a) \partial_t \nu_1(a,t)\, da
	= \int_0^\infty \psi_1(a) \bigg(A_1[\nu_1](a,t)-\lambda \nu_1(a,t)\bigg)\, da\\ 
	&=& \int_0^\infty A_1^+[\psi_1](a) \nu_1(a,t)-\lambda \nu_1(a,t)\, da\\
	&=& \lambda \int_0^\infty \psi_1(a) \nu_1(a,t))\, da - \lambda \int_0^\infty \psi_1(a) \nu_1(a,t))\, da =0.
\end{eqnarray*}
As $x'(t)=0$, we find 
\begin{eqnarray*}
\partial_t\eta_1(a,t) 
&=&
\partial_t\bigg(\nu_1(a,t)-x(t)\varphi_1(a)    \bigg)
= A_1\nu_1(a,t)-\lambda \nu_1(a,t)-0\,\varphi_1(a)\\
&=&
A_1[\eta_1(a,t)+x(t)\varphi_1(a)]-\lambda (\eta_1(a,t)+x(t)\varphi_1(a))\\
&=& A_1[\eta_1](a,t)-\lambda \eta_1(a,t) +x(t) (A_1[\varphi_1]-\lambda\varphi_1(a)) 
= A_1[\eta_1](a,t)-\lambda \eta_1(a,t),\\
\partial_t\eta_2(a,t) 
&=&
\partial_t\bigg(\nu_2(a,t)-(1-x(t))\varphi_2(a)    \bigg)
= A_2\nu_2(a,t)-\lambda \nu_2(a,t)\\
&=&
A_2[\eta_2(a,t)+(1-x(t))\varphi_2(a)]-\lambda (\eta_2(a,t)+(1-x(t))\varphi_2(a))\\
&=&
 A_2[\eta_2](a,t)-\lambda \eta_2(a,t) +(1-x(t)) (A_2[\varphi_2]-\lambda\varphi_2(a)) 
= A_2[\eta_2](a,t)-\lambda \eta_2(a,t).
\end{eqnarray*}
Furthermore, 
\begin{eqnarray*}
\eta_i(0,t) &=& \Pi_i^-[\nu](0,t) = \nu_i(0,t)-\varphi_i(0)\int_0^\infty \psi_i(a) \nu_i(a)\, da\\
&=& \int_0^\infty \beta_i(a)\nu_i(a)\, da 
- \int_0^\infty \beta_i(a)\varphi_i(a)\, da\int_0^\infty \psi_i(a) \nu_i(a)\, da\\
&=& \int_0^\infty \beta_i(a)\bigg(\nu_i(a)\, -\varphi_i(a)\, da\int_0^\infty \psi_i(a') \nu_i(a')\, da'\bigg) da
= \int_0^\infty \beta_i(a)\Pi_i^-[\nu_i](a,t)  da.
\end{eqnarray*}	
\par\qed\par\medskip 
The projector $\Pi_i^+$ projects to the exponentially growing solution, which is the asymptotically attracting solution. As we do not work with the population  size $u_i$ but with the relative frequencies $\nu_i$, the exponentially growing solution is mapped to a constant. We have a line of stationary points parameterized by $x\in[0,1]$. The next theorem shows that this line of stationary points is attracting. This theorem is common knowledge, and basically is a reformulation of the Fundamental Theorem of Demography~\cite{Inaba2017}. Note that we use here the assumption from eqn.~(\ref{expoAssumption}), that $e^{\delta a}\varphi(a)\in L^1$ for a small but positive $\delta$.

\begin{theorem}\label{project1}
There are positive constants $C$, $\delta>0$, such that
$$\|\eta_i\|_{L^1}\leq C e^{-\delta t}$$
 and particularly $\lim_{t\rightarrow\infty} \|\eta_i\|_{L^1}=0$.
\end{theorem}
{\bf Proof: }
Let $B_i(t) = u_i(0,t)$. From the Fundamental Theorem of Demography~\cite[proposition 1.9]{Inaba2017}, we know that 
$$B_i(t) = q_{0,i}\,e^{\lambda t}(1+\eps_i(t))$$
where 
$$ |\eps_i(t)|\leq \tilde C e^{\delta_i t}$$
for some $q_{0,i}>0$, $\tilde C_i>0$, and $\delta_i>0$. Note that we can choose  $\delta_i$ small enough such that $e^{\delta_i a}\varphi(a)$ is still in $L^1$. Furthermore, 
$$u_i(a,t) = B_i(t-a)\,e^{\int_0^a\mu_i(s)\, ds} = q_{0,i}\,e^{\lambda (t-a)}\,(1+\eps_i(t-a)\,)\, e^{-\int_0^a\mu_i(s)\, ds}.$$
Now we simply use the definition of $\nu_i(a,t)$ and $\eta_i(a,t)$ to find 
$$\nu_i(a,t) = \frac{q_{0,i}}{N(0)\,e^{\lambda a}}\,\,e^{-\int_0^a\mu_i(s)\, ds}\,(1+\eps_i(t-a)) =  \frac{q_{0,i}}{N(0)\,c_i}\,\varphi(a)\,(1+\eps_i(t-a)).$$
Therewith, 
$$\Pi_i^+[u_i] =  \frac{q_{0,i}}{N(0)\,c_i}\,\bigg(1+\int_0^\infty \psi_i(a')\,\eps_i(t-a')\, da'\,\bigg)\, \varphi(a)
$$
and 
$$\eta_i(a,t) = \nu_i(a,t)-\Pi_i^+[u_i]
= \frac{q_{0,i}}{N(0)\,c_i}\,\bigg(\eps_i(t-a)\,-\,\int_0^\infty \psi_i(a')\,\eps_i(t-a')\, da'\,\bigg)\,\varphi(a).$$
Since $\psi_i(a)$ is bounded and has a compact support, we have 
$$\bigg|\int_0^\infty \psi_i(a')\,\eps_i(t-a')\, da'\bigg|\leq \int_0^\infty \psi_i(a')\,\tilde C e^{-\delta_i(t-a')}\, da'\leq \hat C e^{-\delta_i t}$$
and since $\varphi_i(a)e^{\delta_i a}\in L^1$, also
$$\|\eps_i(t-a)\varphi(a)\|_{L^1} \leq 
\|\tilde C e^{-\delta_i(t-a')}\varphi(a)\|_{L^1} \leq
\hat C\, e^{-\delta_i t}$$
for some $\hat C>0$, and the result follows.
\par\qed\par\medskip

\section{Weak selection}
Until this point, the two population did not interact at all.
We extend our model by weak selection, that allows for interactions. Basically, we have two possibilities: We either assume that interactions act on the birth rate, or on the death rate (or of course, both). We discuss the modification of the birth rate in detail, and show in the appendix the parallel  theory for the death rate. The birth rates $\beta_i(a)$ are replaced by
$$\beta_i(a)\,(1+\eps g_i(u_1/N, u_2/N),$$
where $\psi_i$ are the adjoint eigenfunctions that we did define in the last section. We assume that
$$ g_i: L^1\times L^1\rightarrow \R$$
such that $g_i(.,.)$ are numbers and do not depend explicitly on age. However, these functions ``see'' the full age structure of their arguments. $\eps>0$ is positive but small, and expresses that the rates are only slightly modified, which is the definition of weak selection.\\
Our starting point for weak selection of the birth rates thus reads
\begin{eqnarray}
	(\partial_t+\partial_a)u_i(a,t) &=& -\mu_i(a) \, u_i(a,t)\label{modelSela}\\
	u_i(0,t) &=& \int_0^\infty \beta_i(a)(1+\varepsilon g_i(u_1/N,u_2/N))\, u_i(a,t)\, da\label{modelSelb}\\
	N(t) &=& \int_0^\infty \psi_1 u_1(a,t)\,+\, \psi_2 u_2(a,t)\, da
\end{eqnarray}
Here, $\psi_i(a)$ are the adjoint eigenfunctions introduced above.
The two equations are now coupled by weak selection. Note that $\nu_i=u_i/N$ are homogeneous of degree zero (if we multiply $u_i$ with $\alpha>0$, $\nu_i$ do not change), such that the model is non-linear but still homogeneous of degree $1$. \par\medskip 

We repeat the same steps as above: First, we disentangle the dynamics of the weighted population size $N(t)$ and relative frequencies $\nu_i=u_i/N$, and then we use the spectral projectors defined above to introduce an appropriate coordinate system which allows to investigate the dynamics of the relative frequencies. In an additional last step, we are able by means of singular perturbation theory to find an approximate one-dimensional equation that characterizes the long term behavior of the model in case of $\varepsilon\ll 1$.

\subsection{Relative frequencies}

We derive equations for the dynamics of $N$ and $\nu_i$. We use $A_i$ in the very same way as defined above, 
$$ A_i u(a) = -\partial_a u(a)-\mu_i(a)u(a),\qquad D(A_i)=\{u\in C^1\,|\,u(0)=\int_0^\infty \beta_i(a)\,u(a)\, da\}.$$
That is, the operators $A_i$ do not acknowledge the weak selection. 
However, we also abuse notation and write $A_i[u_i](a,t)$ to denote  the term $ -\partial_a u_i(a,t)-\mu_i(a)u_i(a,t)$, though it is clear that in general $u_i(.,t)\not\in D(A_i)$.

\begin{lemma}\label{adjunctLemma}
\begin{eqnarray}
	\int_0^\infty \psi_i(a)A_i[u_i](a,t)\, da 
	&=&  \lambda \, \int_0^\infty \psi_i(a)\,u_i(a,t)\, da \\
&&		+ \varepsilon\, \psi_i(0)\,g_i(u_1/N,u_2/N)\,\int_0^\infty \,\beta_i(a)\,u_i(a,t)\, da.\nonumber 
\end{eqnarray}
\label{lemma4}
\end{lemma}
{\bf Proof:}
\begin{eqnarray*}
	&&\int_0^\infty \psi_i(a)A_i[u_i](a,t)\, da=\int_0^\infty \psi_i(a)\bigg(- \partial_a u_i(a,t) - \mu_i(a) u_i(a,t)\bigg)\, da \\
	&=& - \psi_i(a) u_i(a,t)\bigg|_{a=0}^{\infty} + \int_0^{\infty}  u_i(a,t)\bigg( \partial_a  \psi_i(a)- \mu_i(a) \psi_i(a)\bigg)\, da\\
	&=&  \int_0^{\infty}  u_i(a,t)\bigg( \partial_a  \psi_i(a)- \mu_i(a) \psi_i(a)+(\beta_i(a)+\varepsilon \beta_i(a)g_i(u_1/N,u_2/N) \psi_i(0)\bigg)\, da\\
	&=& \int_0^\infty A^+_i[\psi_i](a)u_i(a,t)\, da
	+ \varepsilon\, \psi_i(0)\,g_i(u_1/N,u_2/N)\,\int_0^\infty \,\beta_i(a)\,u_i(a,t)\, da
\end{eqnarray*}
\par\qed\par\medskip 
By means of this lemma it is straight to obtain the dynamics of  the new representation $N(t)$, $\nu_1(a,t)$, $\nu_2(a,t)$ of  the original solutions $u_1(a,t)$, $u_2(a,t)$. 
\begin{theorem}
\begin{eqnarray}
	N'(t) &=& \bigg(\lambda 
+ \varepsilon\, \psi_1(0)\,g_1(\nu_1,\nu_2)\,\int_0^\infty \,\beta_1(a)\,\nu_1(a,t)\, da\\
&&	\qquad+ \varepsilon\, \psi_2(0)\,g_2(\nu_1,\nu_2)\,\int_0^\infty \,\beta_2(a)\,\nu_2(a,t)\, da\bigg)\, N(t),\nonumber\\
\partial_t\nu_i(t) &=&  A_i[\nu_i](a,t) -\lambda \nu_i(a,t) \\
&&- 	 \varepsilon\, \psi_1(0)\,g_1(\nu_1,\nu_2)\,\nu_i(a,t)\,\int_0^\infty \,\beta_1(a)\,\nu_1(a,t)\, da\nonumber\\
&&	- \varepsilon\, \psi_2(0)\,g_2(\nu_1,\nu_2)\,\nu_i(a,t)\,\int_0^\infty \,\beta_2(a)\,\nu_2(a,t)\, da,\nonumber\\
\nu_i(0,t) &=& 
 \int_0^\infty \beta_i(a)\, \nu_i(a,t)\, da
 + \varepsilon g_i(\nu_1,\nu_2)\,
  \int_0^\infty \beta_i(a)\, \nu_i(a,t)\, da.
\end{eqnarray}
\end{theorem}
{\bf Proof: }
\begin{eqnarray*}
N'(t) &=& \int_0^\infty \psi_1(a)\partial_tu_1(a,t)+\psi_2(a)\partial_tu_2(a,t)\,da 
= \int_0^\infty \psi_1(a)\,A_1[u_1](a,t)+\psi_2(a)\,A_2[u_2](a,t)\,da\\
	&=& \bigg(\lambda 
	+ \varepsilon\, \psi_1(0)\,g_1(\nu_1,\nu_2)\,\int_0^\infty \,\beta_1(a)\,\nu_1(a,t)\, da
+ \varepsilon\, \psi_2(0)\,g_2(\nu_1,\nu_2)\,\int_0^\infty \,\beta_2(a)\,\nu_2(a,t)\, da\bigg)\, N(t).
\end{eqnarray*}
Furthermore, 
\begin{eqnarray*}
\partial_t\nu_i(t) &=& \frac{\partial_tu_i(a,t)}{N(t)}
-\frac{u_i(a,t)}{N(t)}\,
 \frac{N'(t)}{N(t)}\\
&=& A_i[\nu_i](a,t) -\lambda \nu_i(a,t) 
- 	 \varepsilon\, \psi_1(0)\,g_1(\nu_1,\nu_2)\,\nu_i(a,t)\,\int_0^\infty \,\beta_1(a)\,\nu_1(a,t)\, da\\
&&	- \varepsilon\, \psi_2(0)\,g_2(\nu_1,\nu_2)\,\nu_i(a,t)\,\int_0^\infty \,\beta_2(a)\,\nu_2(a,t)\, da.
\end{eqnarray*}

\begin{rem}
Note that, by the definition $\nu_i(a,t)=u_i(a,t)/N(t)$, and $N(t)=\int_0^\infty\psi_1(a)u_1(a,t)+\psi_2(a)u_2(a,t)\,da$ 
we still have
$$ \|\psi_1\,\nu_1\|_{L^1}+\|\psi_2\,\nu_2\|_{L^1} = 1$$
irrespective of the weak selection terms.
\end{rem}

\subsection{Spectral projectors}
We define, as above, 
\begin{eqnarray}
x(t) &:=& \int_0^\infty \psi_1(a)\,\nu_1(a,t)\, da\\
\eta_i(a,t) &:=& \Pi_i^-[\nu_i](a,t)
\end{eqnarray}
such that  (using $ \|\psi_2\,\nu_2\|_{L^1} = 1-\|\psi_1\,\nu_1\|_{L^1}=1-x(t)$)
$$ 
\eta_1(a,t) = \nu_1(a,t)-x(t)\varphi_1(a),\qquad
\eta_2(a,t) = \nu_2(a,t)-(1-x(t))\varphi_2(a).
$$
\begin{theorem}\label{project2}
	With $\nu_1(a,t)=\eta_1(a,t) +x(t)\varphi_1(a)$ and $\nu_2(a,t)=\eta_2(a,t) +(1-x(t))\varphi_2(a)$ we have 
\begin{eqnarray}
x'(t) &=&  \varepsilon\,\bigg\{ (1-x(t))\,\,\psi_1(0)\,g_1(\nu_1,\nu_2)\,\int_0^\infty \,\beta_1(a)\,\nu_1(a,t)\, da \\
&&  \qquad -\,x(t)\,\quad \psi_2(0)\,g_2(\nu_1,\nu_2)\,\int_0^\infty \,\beta_2(a)\,\nu_2(a,t)\, da\bigg\}\nonumber\\
\partial_t\eta_i(a,t) &=&  A_i[\eta_i](a,t) -\lambda \eta_i(a,t)+{\cal O}(\varepsilon)\\
\eta_i(a,0) &=& \int_0^\infty \beta_1(a)\eta_1(a,t)\, da+{\cal O}(\varepsilon).
\end{eqnarray}
\end{theorem}
{\bf Proof: }
\begin{eqnarray*}
x'(t)&=& \int_0^\infty \psi_1(a)\,\partial_t\nu_1(a,t)\, da\\
&=& \int_0^\infty \psi_1(a)\,\bigg(
 A_1[\nu_i](a,t) -\lambda \nu_1(a,t) \\
&&\qquad\qquad\qquad - 	 \varepsilon\, \psi_1(0)\,g_1(\nu_1,\nu_2)\,\nu_1(a,t)\,\int_0^\infty \,\beta_1(a')\,\nu_1(a',t)\, da'\nonumber\\
&&\qquad\qquad\qquad - \varepsilon\, \psi_2(0)\,g_2(\nu_1,\nu_2)\,\nu_1(a,t)\,\int_0^\infty \,\beta_2(a')\,\nu_2(a',t)\, da'
\bigg)\, da
\end{eqnarray*}
\begin{eqnarray*}
&=& \int_0^\infty \psi_1(a)\,\bigg(
A_1[\nu_i](a,t) -\lambda \nu_1(a,t)\bigg) da \\
&& - 	 \varepsilon\,x(t)\,\bigg( \psi_1(0)\,g_1(\nu_1,\nu_2)\,\int_0^\infty \,\beta_1(a)\,\nu_1(a,t)\, da
+ \psi_2(0)\,g_2(\nu_1,\nu_2)\,\int_0^\infty \,\beta_2(a)\,\nu_2(a,t)\, da\bigg)\\
&=& \varepsilon\, \psi_1(0)\,g_1(\nu_1,\nu_2)\,\int_0^\infty \,\beta_1(a)\,\nu_1(a,t)\, da \\
&& - 	 \varepsilon\,x(t)\,\bigg( \psi_1(0)\,g_1(\nu_1,\nu_2)\,\int_0^\infty \,\beta_1(a)\,\nu_1(a,t)\, da
+ \psi_2(0)\,g_2(\nu_1,\nu_2)\,\int_0^\infty \,\beta_2(a)\,\nu_2(a,t)\, da\bigg)
\end{eqnarray*}
which yields the result for $x'(t)$. 
\begin{eqnarray*}
&&\partial_t\eta_1(a,t) =
\partial_t\nu_1(a,t)-x'(t)\varphi_1(a)\\
&=&  A_1[\nu_1](a,t) -\lambda \nu_1(a,t) \\
&&- 	 \varepsilon\, \psi_1(0)\,g_1(\nu_1,\nu_2)\,\nu_1(a,t)\,\int_0^\infty \,\beta_1(a)\,\nu_1(a,t)\, da
	- \varepsilon\, \psi_2(0)\,g_2(\nu_1,\nu_2)\,\nu_1(a,t)\,\int_0^\infty \,\beta_2(a)\,\nu_2(a,t)\, da\\
&&-\varepsilon\,\bigg\{ (1-x(t))\,\,\psi_1(0)\,g_1(\nu_1,\nu_2)\,\int_0^\infty \,\beta_1(a)\,\nu_1(a,t)\, da \\
&&  \qquad -\,x(t)\,\quad  \psi_2(0)\,g_2(\nu_1,\nu_2)\,\int_0^\infty \,\beta_2(a)\,\nu_2(a,t)\, da\bigg\}\varphi_1(a)\\
&=&  A_1[\eta_1(a,t) +x(t)\varphi_1(a)](a,t) -\lambda (\eta_1(a,t) +x(t)\varphi_1(a)) +{\cal O}(\varepsilon)\\
&=&  A_1[\eta_1(a,t)](a,t) -\lambda \eta_1(a,t)\varphi_1(a))+{\cal O}(\varepsilon).
\end{eqnarray*}
Similarly, for $\eta_2(a,t)$. Furthermore,
\begin{eqnarray*}
\eta_1(0,t) &=& \nu_1(0,t)-x(t)\varphi_1(0) 
= \int_0^\infty \beta_1(a)\nu_1(a,t)\, da-x(t)\int_0^\infty \beta_1(a)\varphi_1(a)\,da+{\cal O}(\varepsilon)\\
&=& \int_0^\infty \beta_1(a)\eta_1(a,t)\, da+{\cal O}(\varepsilon).
\end{eqnarray*}
Similarly for $\eta_2(a,t)$.
\par\qed\par\medskip 

The next corollary is an immediate consequence of theorem~\ref{project1} and
theorem~\ref{project2}.

\begin{cor}\label{projectCor}
After an initial time layer we find that that
$\|\eta_i(.,t)\|_{L^1}={\cal O}(\eps)$.
\end{cor}

\subsection{Singular perturbation theory and replicator equation}

We first determine the slow manifold. Thereto, we take $\varepsilon$ to zero, and in that freeze the variable $x(t)$. Furthermore, of we have $\varepsilon=0$, the equations for $\eta_i(a,t)$ become 
\begin{eqnarray*}
\partial_t\eta_i(a,t) &=&  A_i[\eta_i](a,t) -\lambda \eta_i(a,t)\\
\eta_i(a,0) &=& \int_0^\infty \beta_1(a)\eta_1(a,t)\, da,
\end{eqnarray*}
such that $\eta_i(a,t)\rightarrow 0$ exponentially fast, and the slow manifold can be written as 
$$ \nu_1(a,t) = x\,\varphi_1(a),\qquad \nu_2(a,t) = (1-x)\varphi_2(a).$$
Due to corollary~\ref{projectCor}, we have also for the perturbed system (after an initial time layer)
$$
\nu_1(a,t) = x\,\varphi_1(a)+{\cal O}(\eps),\qquad \nu_2(a,t) = (1-x)\varphi_2(a)+{\cal O}(\eps).
$$
That is, the system will tend (in lowest order) to the equilibrium situation we obtained for the model without weak selection. On this slow manifold, we find a slow drift induced by weak selection. With eqn.~(\ref{normEverytingTogether})
for the constants appearing in the system, we obtain the following generalized replicator equation.
Let $\tau =\varepsilon t$. The dynamics on the slow manifold in lowest order of $\varepsilon$ is given  by 
\begin{eqnarray}
\frac  d{d\tau} x 
 &=&  x\,(1-x)  \,\bigg\{ \,\,\frac {g_1(x\,\varphi_1,(1-x)\varphi_2)} {\ol A_1}\,
- \,
\frac {g_2(x\,\varphi_1,(1-x)\varphi_2)} {\ol A_2}\,\, \bigg\}
\end{eqnarray}
where, as introduced above,
$$ \ol A_i = 
\frac{\int_0^\infty \,a\, e^{-\int_0^a \mu_i(s)\, ds+\lambda_i\, a} \beta_i(a)\, da}
{\int_0^\infty\,\,\,\,\,  e^{-\int_0^a \mu_i(s)\, ds+\lambda_i\, a} \beta_i(a)\, da}.$$
\par\medskip 
We now proceed to the generalization, that also modifies death weakly, 
\begin{eqnarray}
	(\partial_t+\partial_a)u_i(a,t) &=& -\mu_i(a) [1- \varepsilon m_i(u_1/N,u_2/N) ] \, u_i(a,t)\label{modelSela}\\
	u_i(0,t) &=& \int_0^\infty \beta_i(a)(1+\varepsilon g_i(u_1/N,u_2/N))\, u_i(a,t)\, da\label{modelSelb}\\
	N(t) &=& \int_0^\infty \psi_1 u_1(a,t)\,+\, \psi_2 u_2(a,t)\, da
\end{eqnarray} 
Like for $g_i$, we assume that
$$ m_i: L^1\times L^1\rightarrow \R.$$
Note that the sign convention is such that positive $m_i$, $g_i$ are in favor of subpopulation~1 and to the detriment of subpopulation~2.
For this equation, we obtain by similar arguments as above (see Appendix~\ref{myAppendix}) the extended theorem which states the generalized replicator equation. 
\begin{theorem}
Let $\tau =\varepsilon t$ and $\varphi_i(a)$ the equilibrium age structure of population $i\in\{1,2\}$. The dynamics on the slow manifold in lowest order of $\varepsilon$ is given  by 
\begin{eqnarray}
x' &=& 
 x\,(1-x)  \,
 \bigg\{ \,\,
\quad g_1(x\,\varphi_1,(1-x)\varphi_2)\,/\,\ol A_1\,
- \,
g_2(x\,\varphi_1,(1-x)\varphi_2)\,/\,\ol A_2\,\, \\
&&\qquad\quad\qquad+\,\,\,\,m_1(x\,\varphi_1,(1-x)\varphi_2)\,/\,\ol M_1
-\,m_2(x\,\varphi_1,(1-x)\varphi_2)\,/\,\ol M_2
\bigg\}\nonumber
\end{eqnarray}
where
$$ \ol M_i^{-1} = \int_0^\infty \mu_i(a)\varphi_i(a)\, da = \frac{\int_0^\infty \mu_i(a) e^{-\lambda_i a}l_i(a)\, da}{\int_0^\infty  e^{-\lambda_i a}l_i(a)\, da}
$$
denotes the  life span, averaged in an appropriate sense,  and 
$$ \ol A_i = 
\frac{\int_0^\infty \,a\, e^{-\int_0^a \mu_i(s)\, ds+\lambda_i\, a} \beta_i(a)\, da}
{\int_0^\infty\,\,\,\,\,  e^{-\int_0^a \mu_i(s)\, ds+\lambda_i\, a} \beta_i(a)\, da}\,$$
is the average age at childbearing, both during the exponentially population growth.
\end{theorem}

If not only the exponential growth rates $\lambda_1=\lambda_2$ are identical, but also the rate functions $\beta_1(a)=\beta_2(a)$ and $\mu_1(a)=\mu_2(a)$,
then $\ol A_1=\ol A_2$ resp.\ $\ol M_1=\ol M_2$, and we basically get back the standard replicator equation. However, as neutrality in our sense can be obtained by different parameter functions, the replicator equation is modified. A coevolution of the life-history trait and some other trait might lead to new effects as we will find out in the next section.

\section{Prisoners Dilemma}

We investigate the classical situation in evolutionary game theory: Type~1 produces a public good,  and 
type~2 only profits from the public good, but does not contribute to it. Let us assume that the costs for the
production is $c$, while the benefit is $b$. Then, in simplest case, we define 
\begin{eqnarray*}
 g_1(x\varphi_1, (1-x)\varphi_2) = -c+b\int_0^\infty x\varphi_1(a)\, da  =-c+b\,x,
\quad
g_2(x\varphi_1, (1-x)\varphi_2) = b\int_0^\infty x\varphi_1(a)\, da ) = b\, x
\end {eqnarray*}
while the mortality is not affected ($m_1=m_2=0$).
The generalized replicator equation becomes
\begin{eqnarray}
x' &=& x(1-x)\, [\,(-c+bx)/\ol A_1\,-\,bx/\ol A_2] = \frac{x(1-x)}{\ol A_1\,\ol A_2}\, [-c\,\ol A_2+b\,x\,(\ol A_2-\ol A_1)].
\end{eqnarray}
As an immediate result, we obtain the following theorem.
\begin{theorem}
The stationary solution $x=0$ always is locally asymptotically stable, and $x=1$ is locally asymptotically stable if
$$
b\,(\ol A_2-\ol A_1)>c\,\ol A_2.
$$
\end{theorem}
This result is in line with similar considerations in case of quiescence and seed banks~\cite{Sellinger2019,Mueller2022}. Cooperation only ($x=1$) 
can be stabilized if $b>c$ and
$$ \ol A_2 > \ol A_1\,\,\frac{b}{b-c}$$
but cannot be stabilized if $b\leq c$. We focus on $b>c$.  Cooperation becomes a strategy that cannot be invaded by cheaters, if the 
cheaters have an older average age at childbearing. This observation is a first hint that  coevolution of $\ol A$ and cooperation allows a certain degree of cooperation to become a convergence stable evolutionary stable state. Also that finding parallels the results in~\cite{Sellinger2019,Mueller2022}. \par\medskip 

Thereto, we refine the model, and allow for a certain degree $b$ of public good production for  given type and use the concepts of adaptive dynamics~\cite{Geritz1998} to investigate the dynamics of the degree of cooperation under the pressure of evolutionary forces. The costs will be a non-decreasing function of the costs, that is, is a function $C=C(b)$. Coevolution leads to the fact that also $\ol A$ is a function of $b$, $\ol A=\ol A(b)$. We assume that $C(b)$ and $\ol A(b)$ are smooth (three times differentiable), and in line with the considerations above that $C(b)\leq b$. Furthermore, without cooperation there are also no costs for cooperation, $C(0)=0$.\\
With this notation, we assume that the degree of cooperation for type~1 is $b_1$ (and hence the related costs $c_1=C(b_1)$ and the average childbearing age $\ol A_1 = \ol A(b_1)$), and the degree of cooperation for type~2 is $b_2$  (with $c_2=C(b_2)$ and $\ol A_2 = \ol A(b_2)$). The functions $g_i$ become
\begin{eqnarray*}
g_1(x\varphi_1, (1-x)\varphi_2) &=& -C(b_1) + b_1\int_0^\infty x\varphi_1(a)\, da + b_2\int_0^\infty (1-x)\varphi_2(a)\, da \\
&=& -C(b_1) +b_1\,x+b_2\,(1-x)\\
g_2(x\varphi_1, (1-x)\varphi_2) &=& -C(b_2) + b_1\int_0^\infty x\varphi_1(a)\, da + b_2\int_0^\infty (1-x)\varphi_2(a)\, da \\
&=& -C(b_2) +b_1\,x+b_2\,(1-x).
\end{eqnarray*}
The replicator equation becomes
\begin{eqnarray*}
x'
&=&
\frac{x(1-x)}{\ol A(b_1)\ol A(b_2)}\,{\cal G}(x;b_1,b_2) \\
{\cal G}(x;b_1,b_2) &=& [ -C(b_1) +b_1\,x+b_2\,(1-x)]\,\ol A(b_2)-[ -C(b_2) +b_1\,x+b_2\,(1-x)]\,\ol A(b_1).
\end{eqnarray*}
Let us assume that type~2 is the resident, and type~1 is a rare mutant. We want to understand under which conditions the rare mutant with a degree of cooperation slightly larger but close to that of the resident is able to invade. This is the case if ${\cal G}(x;b_1,b_2)>0$ for $x\in(0,1)$. We assume that $b_1>b_2$, but 
$|b_1-b_2|\ll 1$. As
\begin{eqnarray*}
D(b_2) := \frac{\partial}{\partial b_1} {\cal G}(x;b_1,b_2)\bigg|_{b_1=b_2} 
&=&(C(b_2)-b_2)\, \ol A'(b_2)  -C'(b_2)\ol A(b_2) 
\end{eqnarray*}
is independent on $x$, the function ${\cal G}(x;b_1,b_2)$ will not change sign if $x\in[0,1]$ and $|b_1-b_2|$ sufficiently small and 
$D(b_2) \not = 0$. Particularly,  ${\cal G}(x;b_1,b_2)>0$ for $|b_1-b_2|$ sufficiently small 
and $x\in[0,1]$ if and only if  
\begin{eqnarray}\label{inequ}
D(b_2)>0\quad\Leftrightarrow\quad (C(b_2)-b_2)\, \ol A'(b_2)  > C'(b_2)\ol A(b_2).
\end{eqnarray}
This is the central result to use adaptive dynamics for the analysis and (\ref{inequ}) characterizes the situation when the degree of cooperation will increase by the assumptions of adaptive dynamics. \par\medskip 
If the resident is not cooperating ($b_2=0$), and the costs are strongly increasing in the sense that $C'(0)>0$, cooperation cannot invade. 
In this case, $D(0)<0$, and the strategy $b_2=0$ forms a convergent stable evolutionary stationary strategy, that is, an evolutionary stable strategy.\par\medskip 
If, however, $C'(0)=0$, we have $D(0)=0$. The strategy $b_2=0$ is a convergent unstable evolutionary stationary strategy if $D'(0)>0$,
that is, if 
$$ (C'(0)-1)A'(0)-C'(0)A'(0)-C''(0)A(0)>0 
\quad\Leftrightarrow\quad 
-A'(0) > C''(0)A(0).
$$
As $C$ is non-decreasing, $C''(0)\geq 0$. If $A(b)$ is decreasing fast enough, cooperation can spread; it is necessary that there is a minimal amount of cooperation (which might be introduced by stochastic effects due to a finite population size), then the degree of cooperation will increase. 

\begin{cor} If $C'(0)>0$, then the strategy ``no cooperation''' ($b=0$) is a convergent stable evolutionary stationary strategy.\\
If $C'(0)=0$, and $A'(0)$ sufficiently negative such that 
$$A'(0) < -C''(0)A(0),$$
the strategy ``no cooperation'' ($b=0$) is an evolutionary stationary strategy but convergent unstable; cooperation can spread in this case.
\end{cor}

\section{Discussion}

In this study, we consider an age structured population for frequency dependent selection. We treat the age as a continuous variable, which reduces the complexity of previous matrix approaches~\cite{Li2015,Argasinski2021,Lessard2018}. 
We solely focus on the  weak selection regime, where we can obtain a generalized replicator equation which augments the standard replicator equation by equilibrium properties of the age structure of the population.\\ 
The derivation of this equation  is mainly based on the appearance three different time scales. Interestingly enough, in the context of social and economic sciences, these three relevant time scales have been described earlier~\cite[chapter 2~]{Spruyt1996}:  In economy, individuals interact with each other. Behavioral changes at this level happen fast, at a time scale of few years only. 
The next level is social history describing political shifts. These processes require more time, and range in the time scale of 20-50 years. 
On the last and much slower level history itself moves on. For example, village life changes from Middle Ages to early twentieth century.  We find back these three time scales during the mathematical analysis of our model: An individual responds and interacts with other individuals on the fastest time scale. This time scale is below or at the life span of an individual. The life-history trait tends within a few generations into some equilibrium (e.g., the population reaches the vicinity of a stable age distribution). The slowest time scale is that of evolution itself. Behavioral patterns change, driven by evolutionary forces and mutations on genome level.\\ 
We claim, also in accordance with earlier findings considering age structure~\cite{Argasinski2021,Lessard2018} and  different life-history traits as seedbanks and quiescence~\cite{Sellinger2019,Mueller2022}, that these time scales are generically present in models combining frequency dependent selection and life-history traits, such that the fundamental ideas developed here  will be useful in a wide range of theoretical approaches.\par\medskip 

As a result of the interplay of these three time scales, we find that the interaction of life-history traits and traits described by game theory yield a generalized replicator equation. This generalized replicator equation is as simple as the original one, but included the characteristics of age as a life-history trait into account: The net interaction strength are scaled by the equilibrium age distribution, and the frequency-dependent selection addressing the reproduction is modified  by $\ol A_i$, the average age at reproduction, while that affecting mortality is modified by $\overline M_i$, the average age span. The findings are similar to those in~\cite{Lessard2018}, but more clearly in the interpretation.\par\medskip 

The generalized replicator equation is, under certain conditions, able to produce qualitative new evolutionary stable behavioral pattern which cannot be found in models neglecting  life-history traits. Particularly, by means of adaptive dynamics, it is possible to identify mechanisms that allow for cooperation as an evolutionary stable trait. By means of individual-based simulation models  it has been shown before that age structure can be a mechanism that helps to stabilize cooperation~\cite{Wu2009,Szolnoki2009,Rong2010,Liu2012,Rong2013,Souza2020}, but the mechanism remains rather unclear. The present mathematical analysis reveals the role of time scales and the effect of life-history traits, along the lines of a similar analysis for quiescence and seedbanks~\cite{Sellinger2019,Mueller2022}. Particularly, we find that defection is an invadable strategy, but if costs are small for a weakly cooperating traits, while cooperators  decrease the reproductive age, the strategy ``defect''becomes evolutionary unstable. This finding parallels the idea in~\cite{Krivan2017}, that the timing of interactions can be crucial in frequency-dependent selection. The interaction time is here replaced by the time scale of reproduction, where - we need to emphasize that - the life-history trait {\it per se}, without frequency dependent selection, is purely neutral. Only the interplay between frequency-dependent selection and life-history trait induces crucial fitness differences. In that, cooperation, introduced by small mutations, may spread and enforces itself in the long run, until the level of cooperation tends to an evolutionary stable state. However, this mechanism requires a strong coupling of life-history trait resp.\ reproduction and cooperation, for example by pleyotropy. It remains an open and interesting question  if the hypothesis of co-evolution of cooperation and life-history traits are indeed able to contribute to the solution of the problem how cooperation becomes an evolutionary stable trait. In any case, succesfull reproduction in higher animals obviously is a cooperative task, such that a strong connection cannot be excluded.  
\par\bigskip

{\bf Acknowledgment.}  This research is supported by a grant from the Deutsche Forschungsgemeinschaft (DFG) through TUM International Graduate School of Science and Engineering (IGSSE), GSC 81, within the project GENOMIE QADOP (SJ+JM).

\bibliographystyle{abbrv} 
\bibliography{ageStruct}

\begin{appendix}
\section{Extended model for weak selection}\label{myAppendix}
We extend our model by weak selection on death rate as well, 
\begin{eqnarray}
	(\partial_t+\partial_a)u_i(a,t) &=& -\mu_i(a) [1- \varepsilon m_i(u_1/N,u_2/N) ] \, u_i(a,t)\label{modelSela}\\
	u_i(0,t) &=& \int_0^\infty \beta_i(a)(1+\varepsilon g_i(u_1/N,u_2/N))\, u_i(a,t)\, da\label{modelSelb}\\
	N(t) &=& \int_0^\infty \psi_1 u_1(a,t)\,+\, \psi_2 u_2(a,t)\, da
\end{eqnarray}
We assume that like $g_i$,
$$ m_i: L^1\times L^1\rightarrow \R$$
such that $m_i(.,.)$ are numbers and do not depend explicitly on age. However, these functions ``see'' the full age structure of their arguments.\\


\subsection{Relative frequencies}
Also here, we stick to the definition $A_i[u_i]=-\partial_au(a)-\mu_i(a)u(a)$. Thus, lemma~\ref{adjunctLemma} is unchanged and
$$ \partial_t u_i(a,t) = A_i[u_i](a,t)+\,\varepsilon\,\mu_i(a)\,m_i(u_1/N,u_2/N)\,u_i(a,t).$$
As before we define $\nu_i(t,a) = u_i(t,a)/N(t)$. Therewith we obtain the next theorem. 

\begin{theorem}
\begin{eqnarray}
 N'(t) &=& N(t)\,\bigg[
\lambda  
+ \varepsilon\,\bigg( \sum_{i=1}^2 \psi_i(0)\,g_i(u_1/N,u_2/N)\,\int_0^\infty \,\beta_i(a)\,\nu_i(a,t)\, da\\
&&\qquad\qquad\qquad\qquad\quad+ \,\varepsilon\,m_i(u_1/N,u_2/N)\,\int_0^\infty \mu_i(a)\,\nu_i(a,t)\,da \bigg) \bigg]  \nonumber\\
 \partial_t \nu_i(a,t) &=& A_i[\nu_i](a,t)  -\varepsilon \mu_i(a) m_i(\nu_1,\nu_2) \nu_i(a,t) -\lambda \nu_i(a,t) \nonumber\\
&& - \varepsilon\,\nu_i(a,t)\,\sum_{j=1}^2 \bigg( \psi_j(0)\,g_j(u_1/N,u_2/N)\,\int_0^\infty \,\beta_i(a)\,\nu_j(a,t)\, da\bigg)\\
&& - \varepsilon\,\nu_i(a,t)\,\sum_{j=1}^2 \bigg(\,m_j(u_1/N,u_2/N)\,\int_0^\infty \mu_j(a)\,\nu_j(a,t)\,da \bigg).\nonumber \\
\nu_i(0,t) &=& 
 \int_0^\infty \beta_i(a)\, \nu_i(a,t)\, da
 + \varepsilon g_i(\nu_1,\nu_2)\,
  \int_0^\infty \beta_i(a)\, \nu_i(a,t)\, da.
\end{eqnarray}
\end{theorem}

{\bf Proof: }
\begin{eqnarray*}
N'(t) &=& 
 \sum_{i=1}^2 \bigg(\int_0^\infty \psi_i(a) \partial_t u_i(a,t) da \bigg) \\
&=&
 \sum_{i=1}^2 \bigg(\int_0^\infty \psi_i(a) A_i[u_i](a,t) +\mu_i(a)\,m_i(u_1/N,u_2/N)\,u_i(a,t)\,da \bigg) 
\\ 
&=&
 \sum_{i=1}^2\bigg(\lambda \, \int_0^\infty \psi_i(a)\,u_i(a,t)\, da\, 
+ \varepsilon\, \psi_i(0)\,g_i(u_1/N,u_2/N)\,\int_0^\infty \,\beta_i(a)\,u_i(a,t)\, da\\
&&\qquad\qquad\qquad\qquad\quad+ \,\varepsilon\,m_i(u_1/N,u_2/N)\,\int_0^\infty \mu_i(a)\,u_i(a,t)\,da \bigg) 
\\ 
&=& N(t)\,\bigg[
\lambda  
+ \varepsilon\,\bigg( \sum_{i=1}^2 \psi_i(0)\,g_i(u_1/N,u_2/N)\,\int_0^\infty \,\beta_i(a)\,\nu_i(a,t)\, da\\
&&\qquad\qquad\qquad\qquad\quad+ \,\varepsilon\,m_i(u_1/N,u_2/N)\,\int_0^\infty \mu_i(a)\,\nu_i(a,t)\,da \bigg) \bigg]
\end{eqnarray*}

Furthermore, 
\begin{eqnarray*}
\partial_t\nu_i(t) 
&=& \frac{\partial_tu_i(a,t)}{N(t)} -\frac{u_i(a,t)}{N(t)}\, \frac{N'(t)}{N(t)}\\
&=& A_i[\nu_i](a,t)  -\varepsilon \mu_i(a) m_i(\nu_1,\nu_2) \nu_i(a,t) -\lambda \nu_i(a,t) \\ \nonumber
&& - \varepsilon\,\nu_i(a,t)\,\sum_{j=1}^2 \bigg( \psi_j(0)\,g_j(u_1/N,u_2/N)\,\int_0^\infty \,\beta_j(a)\,\nu_j(a,t)\, da\bigg)\\
&& - \varepsilon\,\nu_i(a,t)\,\sum_{j=1}^2 \bigg(\,m_j(u_1/N,u_2/N)\,\int_0^\infty \mu_j(a)\,\nu_j(a,t)\,da \bigg).
\end{eqnarray*}

\par\qed\par\medskip 
In zero order of $\varepsilon$, we recover the equations derived in Section~\ref{relativeFrequenciesLinear}. The weak selection, however, introduces perturbations of order ${\cal O}(\varepsilon)$.

\subsection{Spectral projectors}
Now recall that we define $x(t) = \int_0^\infty \psi_1(a)\, \nu_1(a,t)\, da$ and again define $\eta_i$ by 
 $\nu_1(a,t)=\eta_1(a,t) +x(t)\varphi_1(a)$ and $\nu_2(a,t)=\eta_2(a,t) +(1-x(t))\varphi_2(a)$.
. With that, we obtain the 
following theorem.
\begin{theorem}
	We obtain
\begin{eqnarray}
x'(t)&=& 
  \varepsilon \bigg\{  (1-x(t))\,\bigg(\psi_1(0)\,g_1(\nu_1,\nu_2)\,\int_0^\infty \,\beta_1(a)\,\nu_1(a,t)\, da
  + m_1( \nu_1,\nu_2)\,\int_0^\infty \mu_1(a)\,\nu_1(a,t)\,da \bigg)\nonumber\\
&&   -\,x(t)\,\bigg(  \psi_2(0)\,g_2(\nu_1,\nu_2)\,\int_0^\infty \,\beta_2(a)\,\nu_2(a,t)\, da
+m_2(\nu_1,\nu_2)\,\int_0^\infty \mu_j(a)\,\nu_2(a,t)\,da\bigg)\bigg\}\\
\partial_t\eta_i(a,t) &=&  A_i[\eta_i](a,t) -\lambda \eta_i(a,t)+{\cal O}(\varepsilon)\\
\eta_i(a,0) &=& \int_0^\infty \beta_1(a)\eta_1(a,t)\, da+{\cal O}(\varepsilon).
\end{eqnarray}
\end{theorem}
{\bf Proof: }
\begin{eqnarray*}
x'(t)&=& \int_0^\infty \psi_1(a)\,\partial_t\nu_1(a,t)\, da\\
&=& \int_0^\infty \psi_1(a)\,\bigg\{A_1[\nu_1](a,t)  -\lambda \nu_1(a,t)
  -\varepsilon \mu_1(a) m_1(\nu_1,\nu_2) \nu_1(a,t) \nonumber\\
&& \qquad\qquad - \varepsilon\,\nu_1(a,t)\,\sum_{j=1}^2 \bigg( \psi_j(0)\,g_j(u_1/N,u_2/N)\,\int_0^\infty \,\beta_i(b)\,\nu_j(b,t)\, db\bigg)\\
&& \qquad\qquad - \varepsilon\,\nu_1(a,t)\,\sum_{j=1}^2 \bigg(\,m_j(u_1/N,u_2/N)\,\int_0^\infty \mu_j(b)\,\nu_j(b,t)\,db \bigg)\,\bigg\}\, da\\
&=& 
  \varepsilon \bigg\{ \psi_1(0) g_1(\nu_1,\nu_2)  \int_0^\infty \beta_1(a)\nu_1(a,t)\,da 
  -  \int_0^\infty \mu_1(a) m_1(\nu_1,\nu_2) \nu_1(a,t)\,da \nonumber\\
&& \qquad\qquad -\,x_1(a)\,\sum_{j=1}^2 \bigg( \psi_j(0)\,g_j(u_1/N,u_2/N)\,\int_0^\infty \,\beta_i(b)\,\nu_j(b,t)\, db\bigg)\\
&& \qquad\qquad - \,x_1(a)\,\sum_{j=1}^2 \bigg(\,m_j(u_1/N,u_2/N)\,\int_0^\infty \mu_j(b)\,\nu_j(b,t)\,db \bigg)\,\bigg\}\\
&=& 
  \varepsilon \bigg\{  (1-x(t))\,\,\psi_1(0)\,g_1(\nu_1,\nu_2)\,\int_0^\infty \,\beta_1(a)\,\nu_1(a,t)\, da \\
&&  \qquad\qquad -\,x(t)\,  \psi_2(0)\,g_2(\nu_1,\nu_2)\,\int_0^\infty \,\beta_2(a)\,\nu_2(a,t)\, da\\
&& \qquad\qquad + \,(1-x_1(a))\,m_1( \nu_1,\nu_2)\,\int_0^\infty \mu_1(b)\,\nu_1(b,t)\,db\\
&& \qquad\qquad - \,x_1(a)  \,m_2(\nu_1,\nu_2)\,\int_0^\infty \mu_j(b)\,\nu_2(b,t)\,db \,\bigg\}
\end{eqnarray*}
The equations for $\eta_i(a,t)$ follow in a similar way as above.
\par\qed\par\medskip

\subsection{Singular perturbation theory and replicator equation}

As above, we use that 
$$ \nu_1(a,t) = x(t)\,\varphi_1(t)+{\cal O}(\eps), \qquad \nu_1(a,t) = (1-x(t))\,\varphi_2(t)+{\cal O}(\eps).$$
Therewith, we obtain (after rescaling time, $\tau=\eps\,t$) the lowest order approximation for $x(t)$,
\begin{eqnarray}
x' &=& 
 x\,(1-x)  \,
 \bigg\{ \,\,
 \frac {g_1(x\,\varphi_1,(1-x)\varphi_2)} {\ol{A_1}}\,
- \,
\frac {g_2(x\,\varphi_1,(1-x)\varphi_2)} {\ol{A_2}}\,\, \\
&&\qquad\qquad\qquad +m_1(x\,\varphi_1,(1-x)\varphi_2)\,\ol \mu_1
-m_2(x\,\varphi_1,(1-x)\varphi_2)\,\ol \mu_2
\bigg\}\nonumber
\end{eqnarray}
where
$$ \ol \mu_i = \int_0^\infty \mu_i(a)\varphi_i(a)\, da = \frac{\int_0^\infty \mu_i(a) e^{-\lambda_i a}l_i(a)\, da}{\int_0^\infty  e^{-\lambda_i a}l_i(a)\, da}
$$
denotes the average death rate in the exponentially growing population.

\end{appendix}

\end{document}